# Search for Millicharged Particles at SLAC

A. A. Prinz, R. Baggs, J. Ballam,* S. Ecklund, C. Fertig, J. A. Jaros, K. Kase, A. Kulikov, W. G. J. Langeveld, R. Leonard, T. Marvin, T. Nakashima, W. R. Nelson, A. Odian, M. Pertsova, G. Putallaz, A. Weinstein
*Stanford Linear Accelerator Center, Stanford, California 94309*

Particles with electric charge $q \leq 10^{-3}e$ and masses in the range 1–100 MeV/$c^2$ are not excluded by present experiments. An experiment uniquely suited to the production and detection of such "millicharged" particles has been carried out at SLAC. This experiment is sensitive to the infrequent excitation and ionization of matter expected from the passage of such a particle. Analysis of the data rules out a region of mass and charge, establishing, for example, a 95%-confidence upper limit on electric charge of $4.1\times10^{-5}e$ for millicharged particles of mass 1 MeV/$c^2$ and $5.8\times10^{-4}e$ for mass 100 MeV/$c^2$.

The quantization of electric charge is an empirically well-supported idea. Of the numerous searches for fractional charge carried out thus far, none has provided conclusive evidence for charge non-quantization. The current bounds on the proton-electron charge difference [1] and the neutron charge [2], of order $10^{-21}e$, lend strong support to the notion that charge quantization is a fundamental principle. However, the Standard Model with three generations of quarks and leptons does not impose charge quantization. In order to enforce quantization of charge, physics beyond the Standard Model is necessary [3]. This observation has stimulated inquiry into mechanisms whereby charge quantization (and perhaps even charge conservation) might be violated [4]. Particles with small fractional charge ($q \lesssim 10^{-2}e$) appear as a natural consequence of many of these mechanisms. There has been interest in the possibility of a small, nonzero electric charge for the neutrino [5], and the possibility that particles with small fractional charge account for a portion of the dark matter in the universe [6]. Additionally, a noteworthy model has been proposed wherein certain particles could exhibit apparent fractional charge without violating charge quantization [7]. Several authors have investigated constraints, imposed by laboratory experiments and by astrophysical and cosmological arguments, on the existence of (free) fractionally charged particles [8]. They point out that there remains a large domain in mass and charge ($10^{-6} \lesssim q/e \lesssim 10^{-3}$, $1 \lesssim M/\frac{\mathrm{MeV}}{c^2} \lesssim 10^4$) where such particles have not yet been excluded. These "millicharged" particles (or "mQ's") could easily escape detection in experiments not specifically designed to observe them.

A dedicated search for mQ's has recently been carried out at SLAC. This search is sensitive to particles with electric charge in the range $10^{-1}$–$10^{-5}e$, and masses between 0.1 and 1000 MeV/$c^2$, whose primary mode of interaction is electromagnetic [9]. The experiment is located near the positron-production target of the SLC, which is well-suited for electromagnetic production of mQ's. The high-intensity, short-duration pulses of the SLC beam allow substantial reduction of backgrounds, as the signal is expected to occur within a narrow window surrounding the arrival time of each pulse. Due to their small electric charge, mQ's traveling through matter interact only rarely. Those with masses greater than about 0.1 MeV/$c^2$ lose energy predominantly through ionization and excitation. To detect them we employ a scintillation counter designed to be sensitive to signals as small as a single scintillation photon (from a single excitation or ionization). The detector is located 110 meters downstream of the positron-production target, with sandstone filling most of the intervening distance (see Fig. 1). Ordinary charged particles (including muons) produced in the target are ranged out in less than 90 meters, leaving, in principle, mQ's as the only beam-related charged particles that reach the detector.

We assume production of mQ's to proceed entirely via electromagnetic interactions. Thus quantum electrodynamics completely characterizes mQ production in terms of the mQ mass, charge ($q$) and spin. The electroproduction cross section, which is proportional to $q^2$, dominates over Bethe-Heitler pair production (which varies as $q^4$) despite being higher order in $\alpha$. We have performed a calculation of the mQ yield expected from the target for spin-1/2 mQ's of various masses, including the effects of showering and scattering of the beam within the target. For mQ's with masses between 0.1 and 100 MeV/$c^2$, the calculations predict a very forward-peaked angular distribution, and indicate that among those mQ's emerging with the smallest angles ($\theta \leq 2$ milliradians), the majority are highly relativistic. It therefore is not essential that the detector cover a large solid angle, and we expect that, in a given pulse, 95% of mQ's reaching our detector (which subtends $2\times2\,\mathrm{mrad}^2$) will arrive within a 1 ns interval. The total yield of mQ's per beam pulse, and the fraction of these that enter the angular acceptance of our detector, are displayed in Table I for four representative values of mQ mass.

The layout of the experiment is shown in Fig. 1. A 29.5-GeV pulsed electron beam from the SLAC linac strikes the 6-radiation-length positron-production target, composed of 75% Tungsten and 25% Rhenium, at a rate of 120 Hz. Each beam pulse contains about $3{\times}10^{10}$ electrons and has a duration of a few picoseconds. Downstream 82.6 meters from the target and 5.3 meters underground is an array of five $21{\times}21$-cm$^2$ scintillation counters which detect high-energy muons produced in the target. These counters monitor the incident-electron flux, determine the beam centroid to verify alignment of the main detector, and fix the arrival time of $\beta \cong 1$ particles in the detector. All other remnants of electromagnetic and hadronic showers produced near the target are absorbed in the stone between the target and the muon counters. The main detector is installed 27.5 meters further downstream in a cylindrical pit, directly in line with the electron beam incident on the target. This is well beyond the range of the most energetic (29.5-GeV) muons. Alignment has been verified to an accuracy of 0.3 milliradians (3 cm) using the position of the muon beam centroid, together with data from a survey of the experiment site carried out after detector installation. The detector consists of a $2 \times 2$ array of blocks of Bicron-408 plastic scintillator, each having dimensions $21{\times}21{\times}130$ cm, and each coupled to an 8-inch hemispherical photomultiplier tube (Thorn EMI model 9353 KA). The longitudinal axis of the array lies along the beam direction.

In order that the sensitivity of the detector extend to pulse heights as small as that of a single photo-electron, steps have been taken to reduce the considerable amount of background noise in this pulse-height region. These include operation of the detector at roughly $0°$ C to reduce thermionic emissions in the tubes, reduction of RF noise using a 0.6-cm thickness of copper shielding, reduction of natural background radiation via a 10-cm thickness of lead shielding, and operation of the tubes at relatively low voltage, with electronic amplification, to further reduce thermionic noise. Additionally, data collection is inhibited for any beam pulse arriving within 30 microseconds of a previous interaction in the detector (in order to minimize the number of phototube after-pulses recorded as events). With the above measures in place, the noise rate per counter is 4 kHz.

Data collection takes place within a 250-ns time gate with leading edge synchronized to pulse passage in the linac. Items recorded include the time (relative to the leading edge) and pulse height of interactions triggering any of the counters in the main detector, plus a signal from a toroid just upstream of the target, which provides a good measure of the number of electrons in the pulse. Similar information is recorded for events in the muon counters. The time distribution for the muon counters is sharply peaked with a FWHM of 2 ns. The presence of mQ's would be indicated by a peak in the time spectrum of the main detector. The expected position of this peak is determined from the observed muon time, a correction for the measured time difference between simultaneously generated events in the muon counters and the main detector (determined using cosmic-ray muons, with the muon counters stacked atop the main detector), the time-of-flight from the muon counters to the main detector for $\beta \cong 1$ particles, and a 15-ns offset to correct for an observed delay between detector triggers due to single scintillation photons (the most probable mQ signal) and those due to traversal by cosmic-ray muons (which generate about 40,000 scintillation photons). To monitor the stability of the timing and verify that the detector is live, an LED mounted on each scintillator is fired at a fixed time within the gate, once every thousand beam pulses. The location of the peak in the time spectrum due to this LED varied less than 2 ns over the duration of the experiment.

We assume that the Bethe-Bloch expression [10] accurately describes the energy loss of mQ's, and that the response of our scintillator is linear with deposited energy, even for energy depositions as small as a single excitation. Calibration of the detector was performed using an $Am^{241}$ source inserted beneath the copper and lead shielding. The most probable energy deposition from the source was estimated via a detailed EGS [11] simulation, and compared to the peak in the measured pulse-height spectrum. The ratio of number of photo-electrons (PE) to deposited energy thus derived is $0.32 \pm 0.03$ PE/keV. A rough check of this value was obtained by repeating the procedure with each of two other sources, $Cs^{137}$ and $Co^{60}$. The stability over time of the calibration value was verified to within 10% by periodic insertion of the $Am^{241}$ source. Using the Bethe-Bloch formula together with this calibration result, one can determine the charge below which a single mQ crossing the detector would generate, on average, less than one PE. This value ($q \cong 3{\times}10^{-3}e$) divides the charge-mass parameter space into two regions. In order to determine the signature of mQ interactions in the detector, one must also divide the charge-mass parameter space according to whether the average number of mQ's entering the detector per beam pulse is greater or less than one. As can be seen from Table I, this division occurs at roughly $Q/M = 4 \times 10^{-4}$, where $Q$ is the charge in units of $e$, and $M$ is expressed in MeV/$c^2$. There are thus four regions to consider. In that for which both the number of incident mQ's per beam pulse ($mQ/pulse$) and the number of PE's per mQ ($PE/mQ$) are greater than one, we would expect a larger-than-minimal pulse height (where "minimal" is that of a single photo-electron, or "SPE") and an event rate of nearly one per beam pulse. The number of mQ events recorded in this case would be far greater than the number of background events. In the region for which $mQ/pulse < 1$ while $PE/mQ > 1$, we again expect a larger-than-minimal pulse height, and, for $M \leq 100\,\mathrm{MeV}/c^2$, a number of recorded mQ events in excess of the number of background events. In the region for which $mQ/pulse > 1$ while $PE/mQ < 1$ there are two possibilities: either a pulse height of minimal (SPE) size and an event rate which could be quite low, or a larger-than-minimal pulse height and an event rate of nearly one per beam pulse (producing, as before, a vast excess of signal events). The last region is that for which both $mQ/pulse$ and $PE/mQ$ are less than one, wherein the expected event rate is low and the expected pulse height is that of an SPE.

The experiment collected data representing a total of $8.4\times10^{18}$ electrons incident on the positron-production target ($2.6\times10^{8}$ beam pulses), over a period of 14 weeks. The pulse-height spectra show a clean SPE peak, and background noise dominated by SPE's. In offline analysis, time information from each of the four counters is corrected for cable length and tube transit-time differences. This information is then combined into a single time spectrum, shown in Fig. 2. The absence of a prominent peak in the time spectrum allows us to immediately rule out the first two regions of charge-mass parameter space discussed in the previous paragraph, plus part of the third. We may safely assume, therefore, that any mQ events detected will be of SPE pulse height. A measurement of the time slewing of SPE-size events in our apparatus leads us to expect a fairly sharp leading edge, long tail and FWHM of 20 ns. We thus choose as our signal region a 40-ns interval surrounding (asymmetrically) the expected arrival time of the mQ signal. Using 40-ns sidebands to the left and right of the signal region as a measure of background, we find a net signal of $207 \pm 382$ events, consistent with zero. (No other 40-ns region in the spectrum shows significant departure from background, so we can rule out the possibility that evidence for mQ's was overlooked due to incorrect identification of signal region.)

A mass-dependent upper limit on the mQ charge is calculated as follows. The expected number of mQ events ($N_{\rm evts}$) is given by

$$N_{\rm evts} = (mQ/pulse)(\Delta E/mQ)\, C\, N_{\rm pulses}\, P_{\rm e}$$

where $mQ/pulse$ is the predicted number of mQ's entering the detector per beam pulse, $\Delta E/mQ$ is the average energy deposited in the detector per incident mQ, $C$ is the detector calibration, $N_{\rm pulses}$ is the total number of beam pulses incident on the positron-production target over the course of the experiment, and $P_{\rm e}$ ($= 0.6\pm0.1$) is a product of efficiencies accounting for deadtime, events lost to the 40-ns time cut, and events with pulse height below discriminator threshold. The uncertainties in the estimated yield of mQ's per pulse (25%) and in $P_{\rm e}$ (17%) are the dominant sources of systematic error. The value of $N_{\rm evts}$, for each of four representative mQ masses, is displayed in the second column of Table II. A 95%-confidence, one-sided upper limit on the number of non-background events ($N_{\rm max}$) is calculated from the measured time spectrum. The limit on mQ charge is then obtained by equating $N_{\rm max}$ to $N_{\rm evts}$. (The actual value we use for $N_{\rm evts}$, to be conservative, is that given in Table II minus its systematic error.) The resulting upper limit on $q$ is presented in the last column of Table II, and the portion of charge-mass parameter space ruled out by this experiment (along with results derived from other experiments) is displayed in Fig. 3. It is worth mentioning that although our analysis incorporates the assumption of linearity of the scintillator for very small energy depositions, an analysis based on the predicted number of mQ-induced delta rays avoids this assumption and results in a limit that is only a factor of 2 less stringent than the one reported in Table II.

In conclusion, a dedicated search for particles with small fractional charge ($q \lesssim 10^{-3}e$) and mass between 0.1 and 1000 MeV/$c^2$ has been carried out downstream of the SLC positron-production target at SLAC [12]. Within the range of charge values to which it was sensitive ($10^{-1}$–$10^{-5}e$), the experiment found no evidence for such particles. The search thus excludes a large region of charge-mass parameter space, significantly improving upon previously established charge limits. The results are summarized in Table II and Fig. 3.

We wish to dedicate this paper to the memory of Joe Ballam. We are grateful to Sacha Davidson and Michael Peskin for informative discussions about millicharged particles, Clive Field for his contributions to the design and construction of the detector, Morris Swartz for his calculations of mQ production rates, and Martin Perl for allowing us the use of his group's resources. We are also grateful for the assistance of Jerry Loomer and our project manager, Glen Tenney. This work was supported by the U. S. Department of Energy under Contract No. DE–AC03–76SF00515.

∗ Deceased.

FIG. 3. Excluded sections of charge-mass parameter space. The dark central region is the area excluded by this experiment. The areas with the lightest shading are those excluded by astrophysical/cosmological arguments, and areas with intermediate shading represent limits derived from other experiments. (Bounds "a" are taken from Davidson *et al.* (1991). Bound "b" is from Golowich and Robinett (1987), and bound "c" is from Davidson and Peskin (1994).)

TABLE I. Calculated yield of mQ's per beam pulse, and fraction produced within the angular acceptance of the detector (for $3\times10^{10}$ incident electrons per pulse). $M$ is the mQ mass in MeV/$c^2$, and $Q$ is the mQ charge in units of $e$.

| $M$ | yield/pulse | fraction accepted |
|---|---|---|
| 0.1 | $(1.55 \pm 0.13)\times10^{9}\,Q^2$ | $0.206 \pm 0.022$ |
| 1 | $(8.51 \pm 0.72)\times10^{7}\,Q^2$ | $0.230 \pm 0.024$ |
| 10 | $(1.56 \pm 0.36)\times10^{6}\,Q^2$ | $0.099 \pm 0.011$ |
| 100 | $(1.90 \pm 0.95)\times10^{4}\,Q^2$ | $0.0414 \pm 0.0039$ |

TABLE II. Predicted number of mQ events in the signal region ($N_{\rm evts}$), and 95%-confidence upper limit on mQ charge ($Q_{\rm max}$) established by this experiment. $M$ is the mQ mass in MeV/$c^2$, and $Q$ is the mQ charge in units of $e$.

| $M$ | $N_{\rm evts}$ | $Q_{\rm max}$ (95% conf.) |
|---|---|---|
| 0.1 | $(6.9 \pm 1.9)\times10^{21}\,Q^4$ | $2.0\times10^{-5}$ |
| 1 | $(4.3 \pm 1.2)\times10^{20}\,Q^4$ | $4.1\times10^{-5}$ |
| 10 | $(3.4 \pm 1.2)\times10^{18}\,Q^4$ | $1.4\times10^{-4}$ |
| 100 | $(1.8 \pm 1.0)\times10^{16}\,Q^4$ | $5.8\times10^{-4}$ |